\documentclass[a4paper]{jpconf}
\usepackage{graphicx}
\begin{document}
\title{Hard probes and the event generator EPOS}

\author{B Guiot and K Werner}

\address{SUBATECH, University of Nantes-IN2P3/CNRS-EMN, Nantes, France}

\ead{guiot@subatech.in2p3.fr}

\begin{abstract}
After a short presentation of the event generator EPOS, we discuss the production of heavy quarks and prompt photons which has been recently implemented. Whereas we have satisfying results for the charm, work on photons is still in progress.
\end{abstract}

\section{Introduction}
Today, there is a large amount of data from p-p, Pb-Pb and recently p-Pb collisions at the LHC which need to be interpreted. EPOS, based on 'Parton-based Gribov-Regge theory' \cite{epo}, aims to reproduce a large range of LHC observables like jets, multiplicity or collective behavior. We will discuss the recent implementation of charm and prompt photons in this event generator. We want hard probes production to be under control for p-p collisions and then, use them for the study of the QGP.\\
First, we will quickly show the general features of EPOS. Then the charm production will be detailed and finally our projet on prompt photons will be exposed.

\section{General presentation}
Some important features of EPOS are :

\vspace{2ex}

\begin{enumerate}
\item Being a real event generator
\item Multiple interactions based on a quantum formalism
\item Perturbative calculation with resummation of collinear corrections at the order $\left(\alpha_s(Q^2)\ln(\frac{Q^2}{\mu^2})\right)^n$
\item Core-corona separation
\item Hydrodynamics done event by event
\item Hadronisation done using a string fragmentation model for the core
\end{enumerate}

\vspace{2ex}

By ``being a real event generator'', we mean that one event in the LHC $\simeq$ one event in EPOS. The program will generate pions even if one is only interested in charm. All particles are registered in a table and, at the end, one has to select particles of interest.\\

\begin{figure}[h]
\includegraphics[width=23pc]{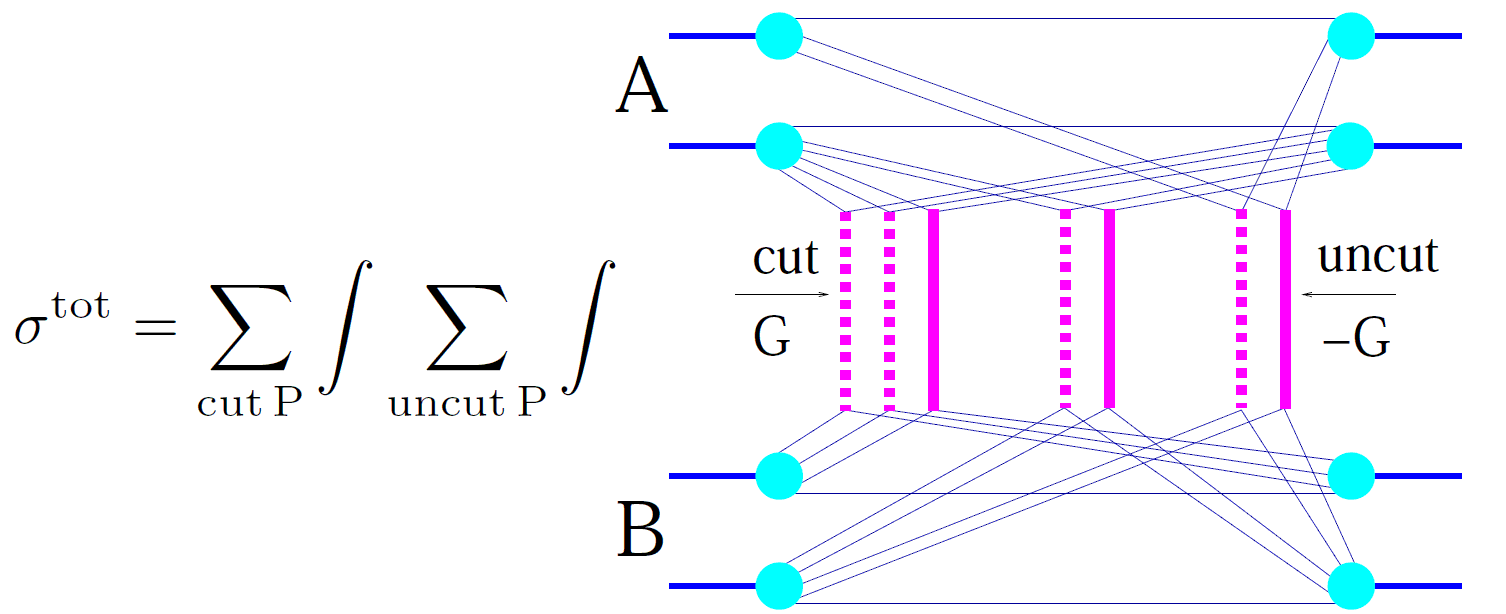}\hspace{2pc}%
\begin{minipage}[b]{14pc}\caption{\label{Tm}Pink lines : pomerons. A and B are nuclei. Small horizontal lines are remnants.}
\end{minipage}
\end{figure}

Our model for multiple interactions is based on a marriage of Gribov-Regge theory \cite{grib,venus} and pQCD. It gives the possibility of a quantum treatment of multiple interactions. By ``based on Gribov-Regge theory'' we mean an assumption about the structure of the $T$ matrix, expressed in terms
of elementary objects called Pomerons (not the same object in EPOS and Gribov-Regge theory). The total cross section can be expressed as illustrated on figure \ref{Tm}.\\
\\
Partial summation provides exclusive cross sections. In Gribov-Regge theory, the elastic amplitude is given by :
\begin{equation}
A_{2\rightarrow 2}(s,t)=\sum_n A_n(s,t)
\end{equation}
with $A_n(s,t)$ corresponding to the amplitude for $n$ pomeron(s) exchange. Following the same idea, we can defined $\sigma_m$ corresponding to the cross section for $m$ cut pomerons. A cut pomeron, figure \ref{pom}, is at the origin of particles production. The multiplicity is then, on the average, proportional to :
\begin{equation}
N\propto m\sum_m \sigma_m
\end{equation}
The treatment is the same for p-p, p-A or A-A collisions.\\

\begin{figure}[h]
\begin{minipage}{18pc}
\includegraphics[width=12pc]{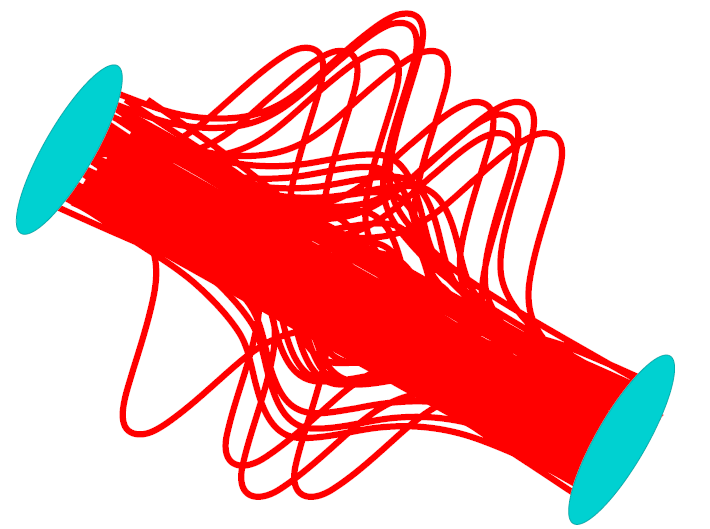}
\caption{\label{flux}Cut pomerons form color flux tubes between the 2 nuclei.}
\end{minipage}\hspace{2pc}%
\begin{minipage}{18pc}
\includegraphics[width=12pc]{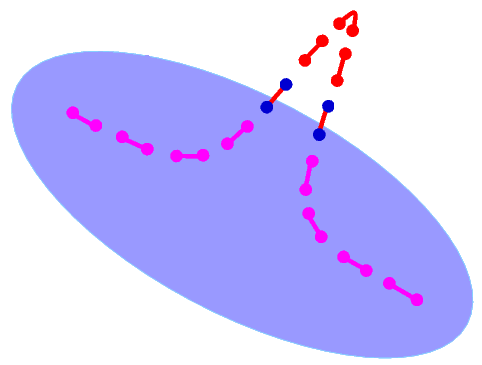}
\caption{\label{jetfluid}Color flux tubes fragment into string piece. The blue region is "matter" formed by high energy density flux tubes.}
\end{minipage} 
\end{figure}

The hydrodynamical evolution is done event by event. Initial conditions are given by the distribution of cut pomerons which correspond to color flux tubes, figure \ref{flux}. Flux tubes fragment into string pieces which will later constitute particles. These flux tubes will constitute both bulk matter (if the energy density is high enough) and jets. ``Matter'' is defined by the region of high energy density flux tubes (the blue region figure \ref{jetfluid}). Then, there are 3 possibilities :
\begin{enumerate}
\item The string piece (the red one) is formed outside the ``matter''. In that case, it simply escapes as a jet.
\item The string piece (the pink one) is formed inside the ``matter'' but has not enough energy to escape. It constitutes the ``matter'' and will evolve with the hydrodynamical code.
\item The string piece (the blue one) is formed inside the ``matter'' and has enough energy to escape (based on energy loss argument). It escapes as a jet which has interacted with the fluid. 
\end{enumerate}
For more details, see \cite{hydro,v2}.
\\

\begin{figure}[h]
\begin{minipage}{18pc}
\includegraphics[width=18pc]{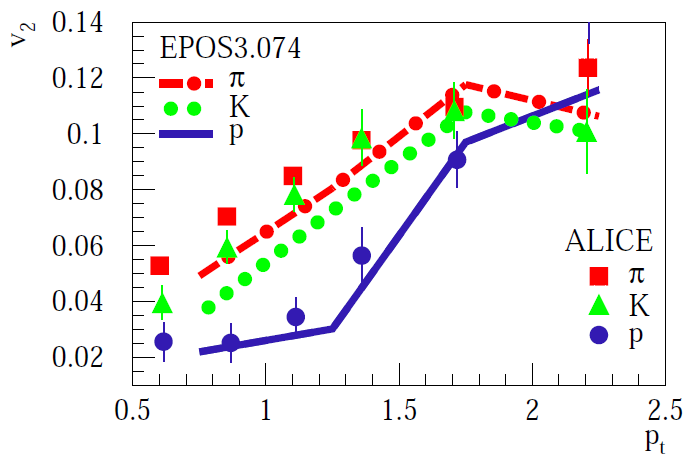}
\caption{\label{v2}(Color online) Elliptical flow coefficients v2 for pi-
ons, kaons, and protons. We show ALICE results (squares)
and EPOS3 simulations (lines). Pions appear red, kaons
green, protons blue.}
\end{minipage}\hspace{2pc}%
\begin{minipage}{18pc}
\includegraphics[width=18pc]{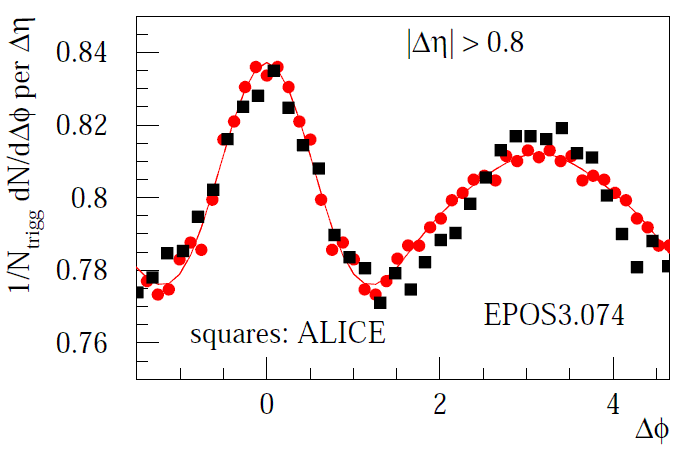}
\caption{\label{ridge}(Color online) Associated yield per trigger, pro-
jected onto $\Delta\phi$, for $|\Delta\eta| >$ 0.8. We show ALICE results
(black squares) and EPOS3 simulations (red dots).}
\end{minipage} 
\end{figure}

 With these prescriptions, we can reproduce the $v_2$ for identified particles or the ridge, even for p-Pb collisions (figure \ref{v2} and figure \ref{ridge}).

\newpage

\section{Charm production}

\begin{figure}[h]
\begin{minipage}{18pc}
\includegraphics[width=18pc]{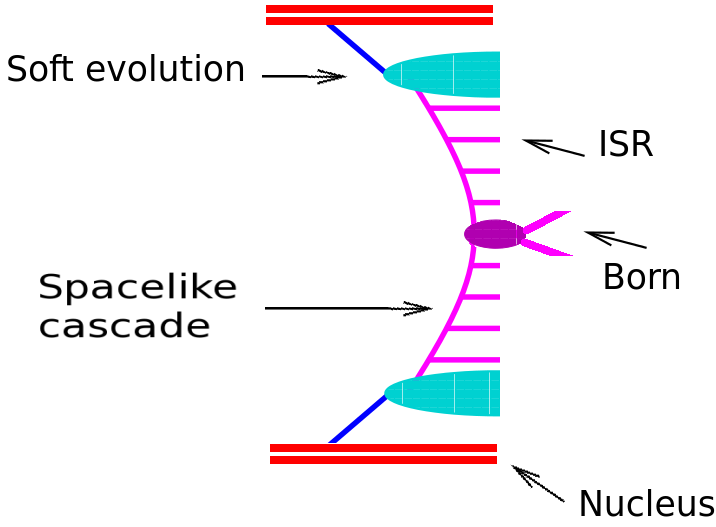}
\caption{\label{pom} A cut pomeron. The center of the pomeron is a (pQCD) ladder diagram.}
\end{minipage}\hspace{2pc}%
\begin{minipage}{18pc}
\includegraphics[width=18pc]{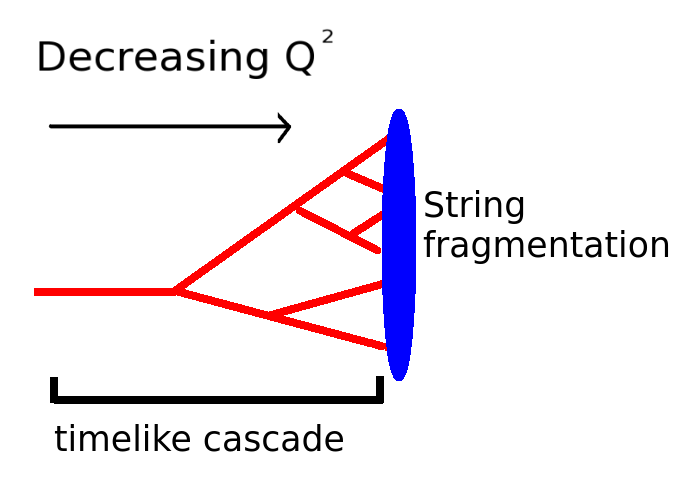}
\caption{\label{tim}Partonic shower and hadronization by string fragmentation.}
\end{minipage} 
\end{figure}

A charm can be produced during the spacelike cascade, the born process, the timelike cascade (partonic shower) and the string fragmentation, see figures \ref{pom} and \ref{tim}. 
Based on DGLAP formalism, spacelike and timelike cascades resumme collinear divergences.\\

During the spacelike cascade, a spacelike parton emits timelike particles until he reaches the born process. In the leading log approximation, virtualities
are strongly ordered :
\begin{equation}
 Q_1^2 \ll Q_2^2 \ll ... \ll Q_{born}^2
\end{equation}
We use the following probability distribution for variables $Q^2$ and $x$ :
\begin{equation}
\frac{dP(Q^2_0,Q^2,x)}{dxdQ^2}\propto \frac{\alpha_s}{2\pi}\frac{p(x)}{Q^2}\Delta(Q^2_0,Q^2) \hspace{1cm} Q_0^2<Q^2 \label{proba1}
\end{equation}
where $p(x)$ are the appropriate splitting functions and $x$ is the fraction of the parent parton light cone momentum $k^+$ :
\begin{equation}
 k^{'+}=xk^+
\end{equation}
$\Delta(Q^2_0,Q^2)$, the Sudakov form factor, gives the probability for no resolvable emissions between $Q^2_0$ and $Q^2$. In text books one can find :
\begin{equation}
r^2=\frac{(1-x)}{x}Q^2-\frac{p_t^2}{x}
\end{equation}
$r$ being the 4-momentum of the emitted (timelike) parton and $p_t$ its transverse momentum. Usually one takes $r^2=0$, but for heavy quarks we choose :
\begin{equation}
r^2=m^2
\end{equation}
which implies :
\begin{equation}
x<\frac{Q^2}{Q^2+m^2}
\end{equation}
The phase space for radiating a heavy quark is smaller than the one for light partons.\\

The born process, in the center of the ladder, is nothing else that the leading order cross sections for $\alpha_s^2$, $\alpha_s\alpha_{el}$ and $\alpha_{el}^2$ processes.\\

Outgoing on-shell partons ($r^2=0$) is an approximation. Out-born partons and those emitted during the spacelike cascade have a finite virtuality :
\begin{equation}
Q^2_{max} \sim p_t^2+m^2
\end{equation}
During the timelike cascade (partonic shower) these partons will loose their virtuality by doing successive splittings. This cascade is stopped 
when $Q^2 \sim \Lambda_{QCD}^2$. The leading log approximation is used and angular ordering \cite{fsr1,fsr2} is implemented.
The emission probability is :
\begin{equation}
\frac{dP(Q^2_0,Q^2,z)}{dxdQ^2}\propto \frac{\alpha_s}{2\pi}\frac{p(z)}{Q^2}\Delta(Q^2_0,Q^2) \hspace{1cm} Q_0^2>Q^2
\end{equation}
$z$ being the splitting variable defined as :
\begin{equation}
z=E_{children}/E_{parent}
\end{equation}

\begin{figure}[h]
\includegraphics[width=21pc]{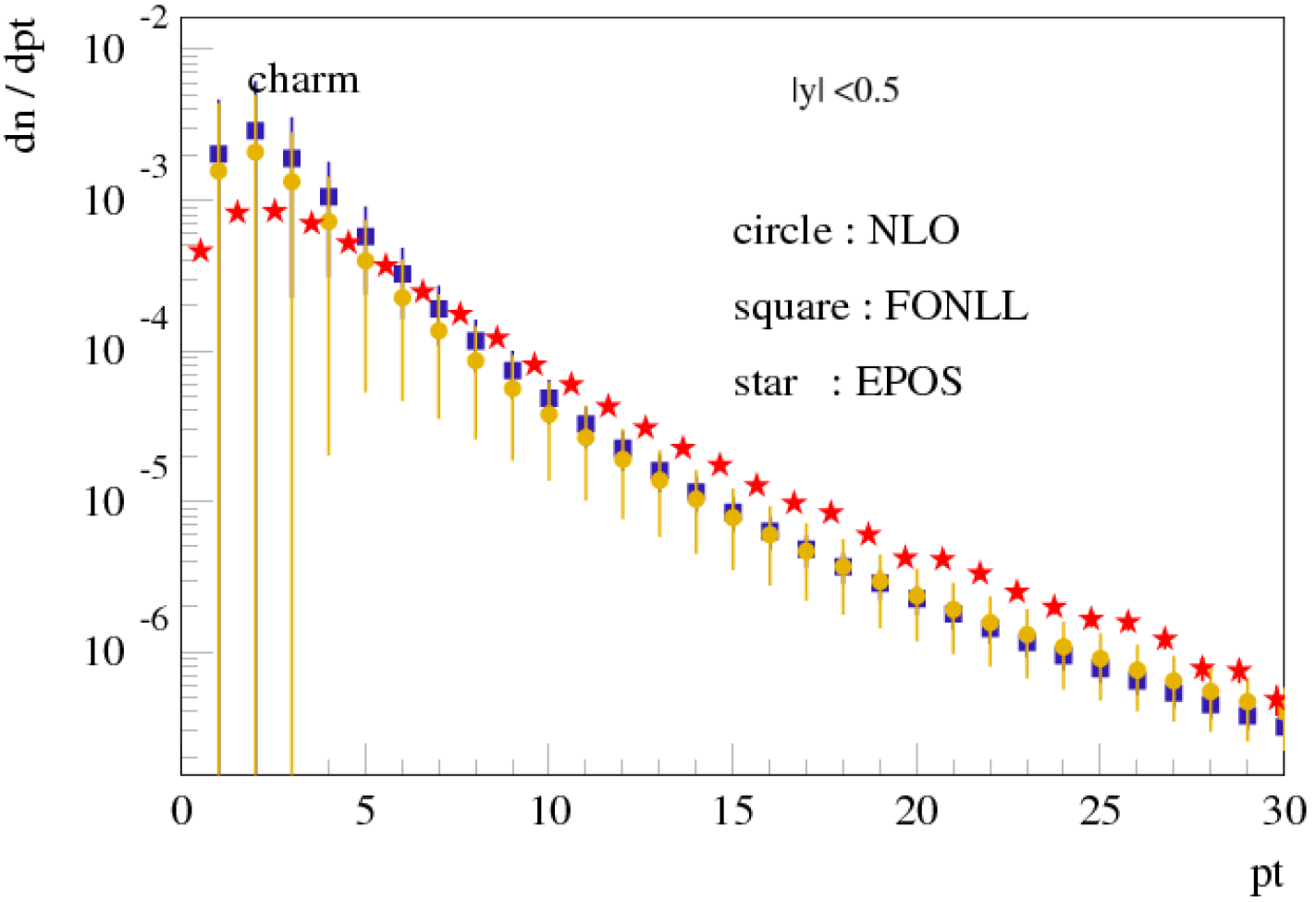}\hspace{2pc}%
\begin{minipage}[b]{14pc}\caption{\label{charm}charm from Cacciari vs EPOS.}
\end{minipage}
\end{figure}

\begin{figure}[h]
\begin{minipage}{18pc}
\includegraphics[width=18pc]{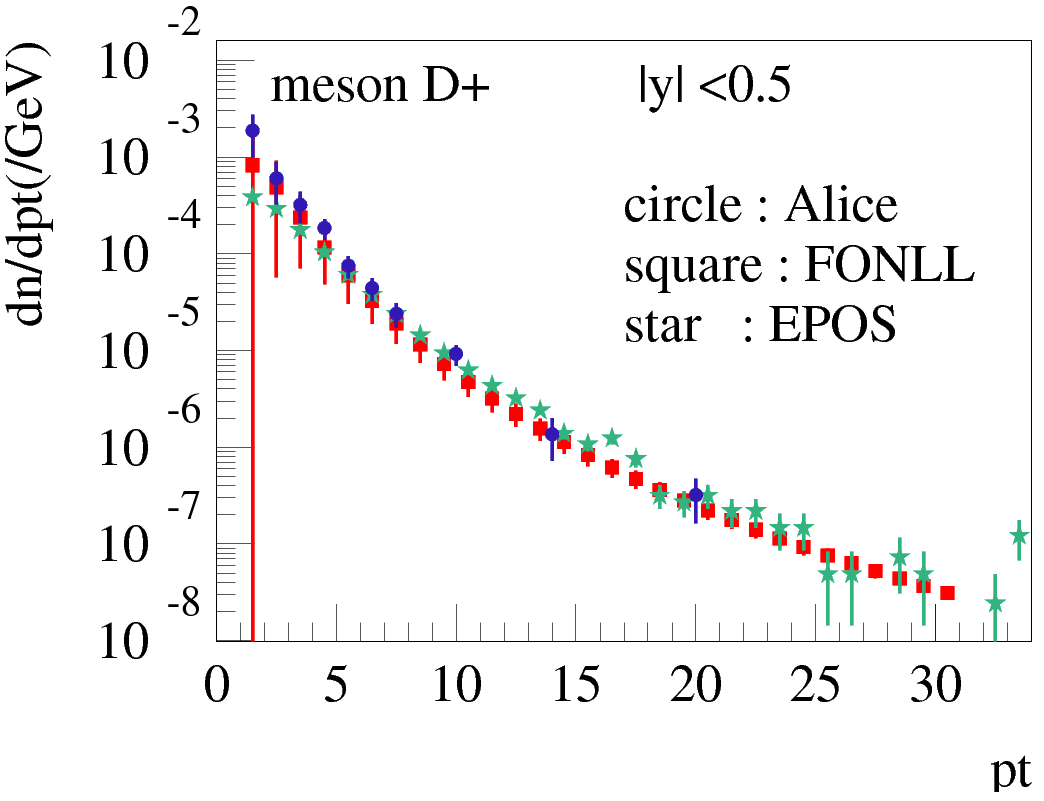}
\caption{\label{D+} D+ yield from EPOS compare to FONLL calculation \cite{fonll} and Alice experiment \cite{alice}.}
\end{minipage}\hspace{2pc}%
\begin{minipage}{18pc}
\includegraphics[width=19pc]{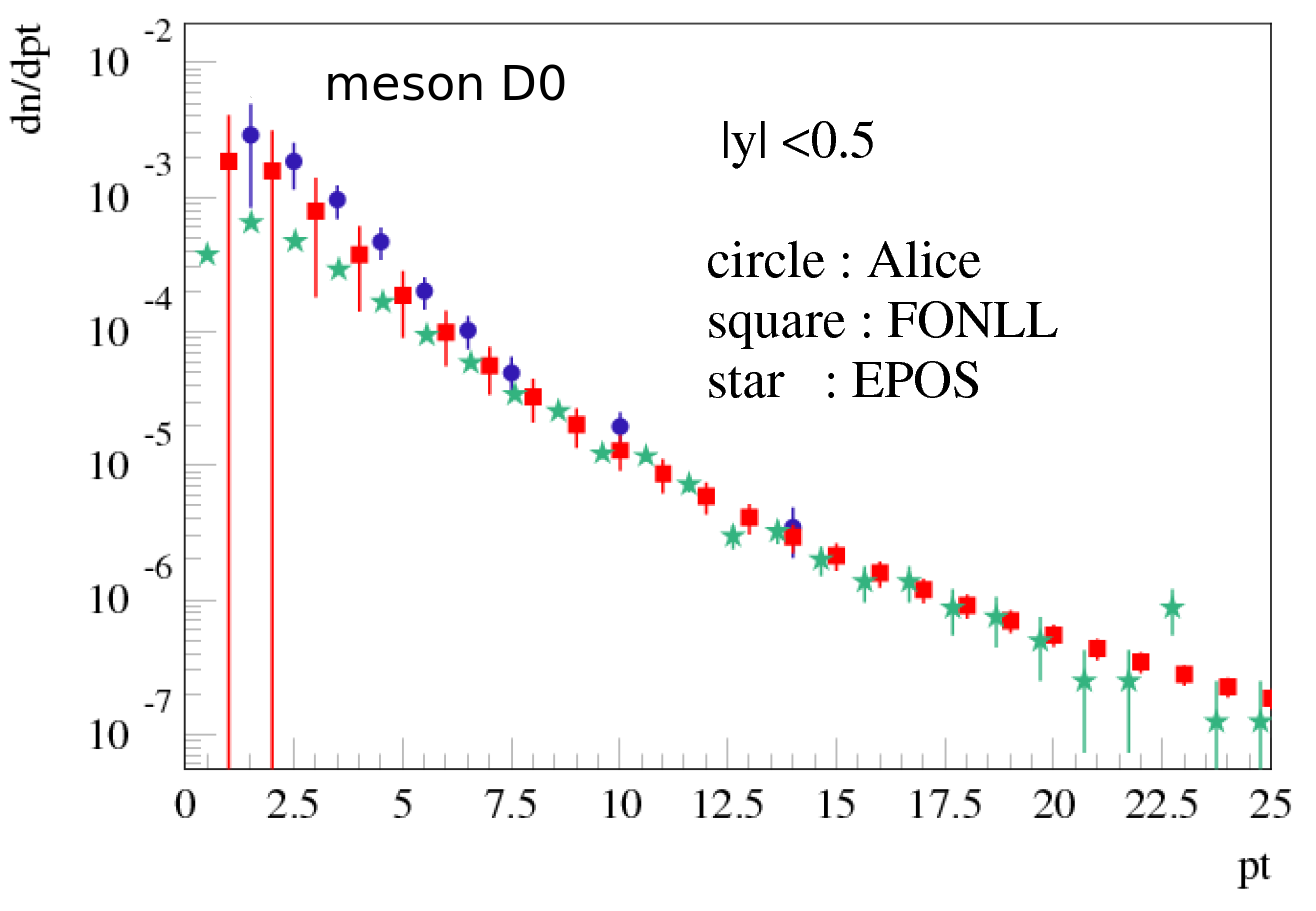}
\caption{\label{D0}D0 yield from EPOS compare to FONLL calculation \cite{fonll} and Alice experiment \cite{alice}.}
\end{minipage} 
\end{figure}

For the implementation of charm production, no parameters have been changed. Our first test is the comparison with FONLL calculations of M. Cacciari \cite{fonll}, 
figure \ref{charm}. At high $p_t$ the shape is the same but our central values are higher.
Next, we test EPOS for $D+$ and $D0$ mesons, by comparing our results with the Alice experiment \cite{alice} and FONLL calculation, figure \ref{D+} and figure \ref{D0}.
\\

\begin{figure}[h]
\includegraphics[width=20pc]{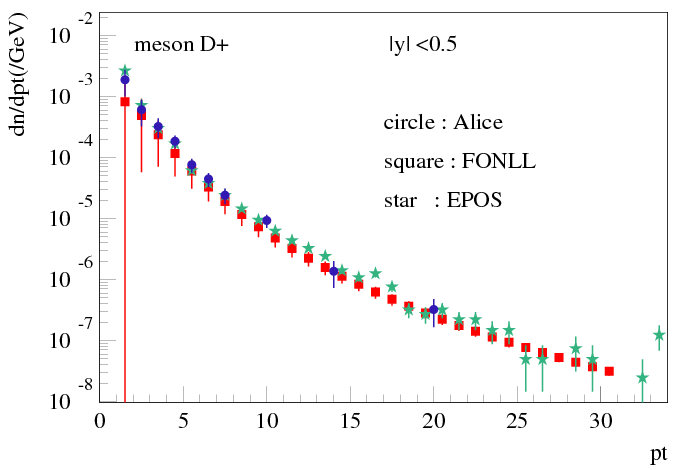}\hspace{2pc}%
\begin{minipage}[b]{14pc}\caption{\label{D+st}D+ from EPOS with $c\bar{c}$ creation during strings fragmentation.}
\end{minipage}
\end{figure}

EPOS uncertainties account only for statistics and there is no charm production during the string fragmentation. Our results are globally in good agreement with Alice and FONLL. 
However, charm quarks are missing at very low $p_t$. The explanation could be that charm production during the timelike cascade is too small. In the game of fitting data, we can reproduce Alice results by allowing charm production in string fragmentation (one parameter is changed), figure \ref{D+st}.

\newpage

\section{Prompt photons}
We want to :
\begin{enumerate}
\item Study isolation criteria
\item Compare EPOS and Jetphox
\item Use EPOS for $\gamma/jet$ and $\gamma/hadron$ correlations
\end{enumerate}
Photons production in the born process was already implemented, but it was not the case for the spacelike and the timelike cascade. 
It has been done with the same formalism used for partons i.e based on the probability eq. \ref{proba1}. 
One needs to replace $\alpha_s$ by $\alpha_{el}$ and use the appropriate splitting function. 
The splitting of a photon into a pair of particle-antiparticle is neglected.\\

Like in experiments, we have an isolation condition on selected photons. Using the table where final particles are registered, 
we define a cone of radius $R$ around the photon candidate. If the sum of transverse energy of particles in this cone is smaller than a given 
value (5 GeV for CMS \cite{cms}), then the photon is isolated. Here, the fact that EPOS is experiment like is important. In Jetphox, the addition of this isolation criteria gives a non-physical rise of the cross section with $1/R$ \cite{jetp}.\\

Work on photons is still in progress. Fragmentation photons are strongly suppressed due to isolation requirement, whereas $\sim 98\%$ of direct photons 
are isolated, table \ref{isolation}.
\begin{center}
\begin{table}[h]
\centering
\caption{\label{isolation}Pourcentage of isolated direct photons as a function of transverse mometum for the isolation criteria used by CMS \cite{cms} ($R=0.4$, $\sum p_t <5$GeV).} 
\begin{tabular}{cc}
\br
$p_t$ & $\#$ of isolated direct photons/$\#$ of direct photons\\
\mr
11 &   0.972\\
13 &   0.981\\
15 &   0.971\\
17 &   0.973\\
  19 & 0.976\\
  21 & 0.992\\
  23 & 0.947\\
  25 & 0.991\\
  27 & 0.976\\
  29 & 0.966\\
  31 & 0.977\\
  33 & 0.978\\
  35 & 0.978\\
\br
\end{tabular}
\end{table}
\end{center}

\begin{figure}[h]
\includegraphics[width=20pc]{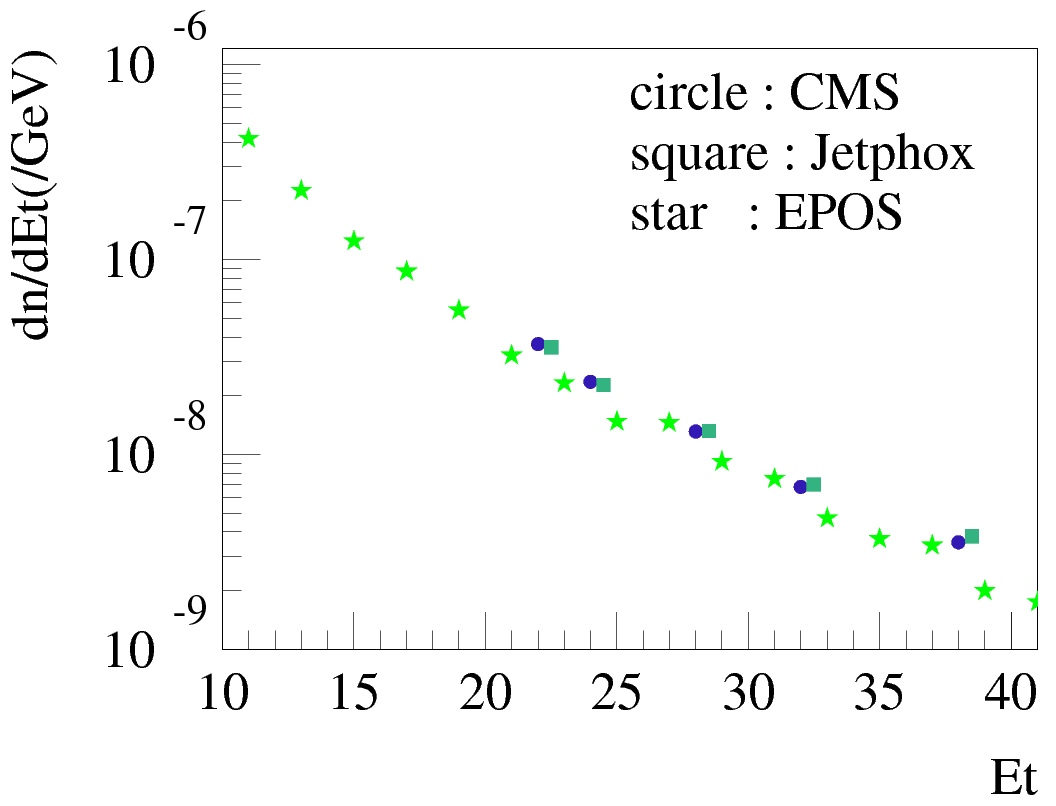}\hspace{2pc}%
\begin{minipage}[b]{14pc}\caption{\label{phot}Isolated photons. For more clarity, a +0.5 shift (x axis) is used for Jetphox data.}
\end{minipage}
\end{figure}

Comparison with CMS \cite{cms}, figure \ref{phot}, shows that our yield seems to be too low by approximately a factor of $1.5$. Comparison with Jetphox will give precious information which, I hope, will allow us to improve our results for photons.

\newpage

\section{Conclusion}
A unified formalism in EPOS is one of its strengths. Heavy quarks and prompt photons production are based on the same equations with the same parameters. Whereas the work on charm is nearly finished, prompt photons physics still need to be improved. We already have very good results for flow \cite{v2,epos3}, and with the implementation of hard probes, EPOS could be an excellent tool for the study of the QGP.

\section*{Acknowledgement}
I would like to thank the \textit{\textbf{projet TOGETHER, region Pays de la Loire}} which has financed my thesis.

\end{document}